\documentstyle[prl,aps,epsfig,multicol]{revtex}
\newcommand{\bm}{\bibitem}

\def\a {\alpha}
\def\b {\beta }
\def\g {\gamma}
\def\G {\Gamma}

\def\O {\Omega}
\def\s {\sigma}
\def\S {\Sigma}
\def\p {\pi}

\def\m {\mu}

\def\pr{\partial}
\def\t {\tau}
\def\v{\vec}
\def\f {\frac}
\def\lt{\left}
\def\rt{\right}

\def\sq{\sqrt}
\def\ra{\rightarrow}

\def\dt{\delta}
\def\Dt{\Delta}
\def\ep{\epsilon}
\def\z {\zeta}

\def\be {\begin{equation}}
\def\ee {\end{equation}}
\def\ba {\begin{array}}
\def\ea {\end{array}}
\def\bea {\begin{eqnarray}}
\def\eea {\end{eqnarray}}

\def\bd {\begin{displaymath}}
\def\ed {\end{displaymath}}
\begin{document}
\title{Isospin mode splitting and mixing  
in asymmetric nuclear matter}

\author{Subhrajyoti Biswas and Abhee K. Dutt-Mazumder}
\address{Saha Institute of Nuclear Physics,
1/AF Bidhannagar, Kolkata-700 064, INDIA}
\medskip

\maketitle
\begin{abstract}
We estimate exclusive density and asymmetry parameter dependent  
dispersion relations of various charged states of pions
in asymmetric nuclear matter. The possibility of matter
induced mixing of $\pi^0$ with $\eta$ is clearly exposed with the
further mass modification of $\pi^0$ meson due to mixing. 
Asymmetry driven mass splitting and 
mixing amplitude are of the same order as  
the corresponding values in vacuum.
Closed form analytic results for the mass shifts and dispersion
relations with and without mixing are presented. Furthermore, we
discuss the sensitivity of our results on the
scalar mean field within the framework of Quantum Hadrodynamics.\\

\leftline{ Keywords : Isospin symmetry, splitting, collective modes}
\end{abstract}
\vspace{0.08 cm}

\section{Introduction}

It is well known that the particle dispersion changes in matter
due to scatterings with the medium constituents. This is characterized
by the density($\rho_b$) dependent self-energy ($\S(q_0,|{\bf q}|))$
of the propagating particle. Clearly it depends on the energy ($q_0$)
and three momentum ($|{\bf q}|$) of the particle. At low density
$\S(q_0,|{\bf q}|)$
can be calculated by multiplying the forward scattering amplitude
with the density, which, however fails at higher density where multiple
scattering becomes important \cite{ericson66}. To incorporate these higher 
order effects
one calculates full self-energy by evaluating loops at various orders.
The real and imaginary parts of $\S(q_0,|{\bf q}|)$ 
determine the in-medium mass and the
decay width of the particle. The medium modified
dispersion relations characterizing the collective excitations 
are obtained from the zeros of the inverse propagator
involving the density dependent self-energies.

In the present paper we plan to investigate the pion propagation in 
asymmetric nuclear matter (ANM).
Such investigations are important to understand the
pion($\pi$)-nucleon($N$)
dynamics at finite density. Medium modifies the pion masses. Such mass shifts in 
nuclear matter can be used to calculate the pion-nucleus optical 
potential \cite{batty97,weise01} which are different for different 
charged states. Furthermore,
in-medium pion dispersion relations also determine the
low mass dilepton yields
in relativistic heavy ion collisions which is a subject of contemporary
debate \cite{helgesson95,rapp00}. With this motivation, we here focus
on the in-medium properties of pions 
in neutron rich matter. This is in
sharp contrast with most of the previous calculations that deal 
with symmetric nuclear matter (SNM) \cite{korpa95,liu97}. 
The possibility of asymmetry driven
$\pi$-$\eta$ mixing, as explored here,
has not been addressed before. Such mixing also modifies the pion mass
in ANM. It should be noted that such mixing of different isospin states
and the mass splitting of various charged states in ANM is a generic
effect and therefore interesting in itself. Furthermore,
in the present study we go beyond Fermi gas 
model and incorporate $N$-$N$ interactions within the framework of Quantum
Hadrodynamics (QHD) \cite{serot86}.

Pion propagation in ANM is qualitatively
different from SNM. This, as we show, is due to
the possible splitting and mixing of the collective modes.
Both the splitting of isospin multiplets and the coupling of the isoscalar
and isovector oscillations are related to the ground state induced isospin 
symmetry breaking \cite{abhee96a,abhee96b,abhee97,abhee01}.
Clearly charged and neutral pion masses will be 
non-degenerate once the density dependent dressings of the pion propagators
in ANM are considered.  Such splitting may also be observed in 
vacuum due to virtual nucleon-antinucleon excitations if
the mass(M) splitting of neutron($n$) and proton($p$) are taken into account 
(or at the quark level $u$-$d$ mass difference). In addition under this
situation $\pi^0$ (pure isospin eigen state) may contain  admixture of 
$\eta$ meson. Such $\pi$-$\eta$ meson mixing in vacuum due to the $n$-$p$ mass
difference has been considered in ref.\cite{piekarewicz93}. But the mechanism 
we propose here for such splitting and mixing is generically different.
Here, as we shall see, it is driven by the difference of the proton
and neutron Fermi momentum ($k^F_{p,n}$), {\em i.e.} $k^F_p\neq
k^F_n$. 
This possibility, in the context of $\rho$-$\omega$ mixing, 
was first suggested by one of the
present authors 
\cite{abhee97}. Moreover, to focus exclusively on the 
density dependent 
effect we neglect explicit
symmetry breaking and  take  $M_p=M_n$.

\section{Formalism}
To describe pion-nucleon interaction we consider pseudovector 
coupling : 
\bea
{\cal L}^{PV}_{int}=-\f{ f_\p}{m_\p}\bar{\Psi}_N\g_5\g_\m\pr^\m
\lt(\v{\t}\cdot\v{\Phi}_\p\rt)\Psi_N
\label{Lpinn}
\eea
Here $\f{f_\p}{m_\p}=\f{g_\p}{2M}$ and $\f{g^2_\p}{4\p}=12.6$.
In Eq.\ref{Lpinn} $\Psi_N$ and $\Phi_\p$ are the nucleon and pion 
fields respectively and $\v\t$ is the isospin 
operator. The $\eta$-$NN$ interaction Lagrangian for pseudovector
coupling can be found from Eq.\ref{Lpinn} with the following
substitution : $\f{f_\p}{m_\p}\ra\f{f_\eta}{m_\eta}$ and 
$\v{\t}\cdot\v{\Phi}_\p\ra \Phi_\eta$. Here 
$\f{f_\eta}{m_\eta}=\f{g_\eta}{2M}$ and $\f{g^2_\eta}{4\p}=5.5$. 

Another essential ingredient to calculate in-medium self-energy 
of the $\pi$/$\eta$ meson is the in-medium nucleon Green's function. 
In nuclear matter usual
vacuum is replaced by the occupied Fermi sea which forbids on
mass-shell nucleon propagation below the Fermi momentum because of the Pauli 
blocking. The relativistic nucleon
propagator in this case is given by $G(k)=G^F(k)+G^D(k)$, where the superscript
$F$ and $D$ denotes the free and dense part respectively. Explicitly 
they are given by
\bea
G^F(k)=\frac{k\!\!\!/+M^*}{k^2-M^{* 2}+i\z},~~~
 G^D(k)=\frac{i\pi(k\!\!\!/+M^*)}{\ep^*}\delta(k_0-\ep^*)
\theta(k^F-|{\bf k}|)
\label{nuclprop}
\eea
Here $k^F$ is the Fermi momentum and $M^*$ 
denotes medium modified nucleon mass. The energy  of nucleon is 
$\ep^*=\sq{M^{*2}+{\bf k}^2}$. We, from now onwards,
use $k_p$ and $k_n$ to denote the
proton and neutron Fermi momentum respectivel. In QHD the effective nuclear mass is determined
from the following self-consistent condition \cite{serot86}.
\bea
M^*=M- \frac{g_\s^2}{M_\s^2} (\rho^s_p+\rho^s_n),
\eea
where $\rho^s_i$ ($i=p,n$) represent scalar densities given by
\bea
\rho^s_i=\frac{M^*}{2\pi^2} \left[\epsilon^*_i k_i - M^{* 2}
\ln \left ( \frac{\epsilon^*_i + k_i}{M^*}\right )\right ]
\label{emass}
\eea

The general formula for calculating self-energy is given below.
\bea
\S_{\p\p}(q) = -~i~\int \f{d^4k}{(2\p)^4}Tr\lt[\{~i \G(q) \} ~iG_{p(n)}(k+q) 
\{~i \G(-q)\} ~iG_{p(n)}(k)  \rt]
\label{self0}
\eea

Where $G_{p(n)}$ denotes the proton (neutron) propagator and 
$\G(q)=-~\f{f_\p}{m_\p}\g_5 i q\!\!\!/$ is the vertex factor. 
With Eq.\ref{nuclprop} and Eq.\ref{self0} the  
pion self-energy due to scattering from Fermi sphere takes the
following form :

\bea
\S_{\p\p}(q)= (-i)\int \f{d^4k}{(2\p)^4}\lt(\f{f_\p}{m_\p}\rt)^2~{\bf T}
\label{self1}
\eea  

 Here {\bf T} is the trace factor. For $\p^0$ 
\bea
{\bf T} = Tr[\g_5q\!\!\!/G^F_p(k+q)\g_5q\!\!\!/G^D_p(k)
 + \g_5q\!\!\!/G^D_p(k+q)\g_5q\!\!\!/G^F_p(k)] + [p \ra n]
\label{self2}
\eea

But for $\p^{+(-)}$,~~$f_\p $ is to be replaced by $\sqrt{2}f_\p$ ($\sqrt{2}$ is the 
isospin factor) and
 
\bea
{\bf T} = Tr[\g_5q\!\!\!/G^F_{p(n)}(k+q)\g_5q\!\!\!/G^D_{n(p)}(k)
 + \g_5q\!\!\!/G^D_{p(n)}(k+q)\g_5q\!\!\!/G^F_{n(p)}(k)]
\label{self3}
\eea

\begin{figure}[htb]
\begin{center}
\epsfig{figure=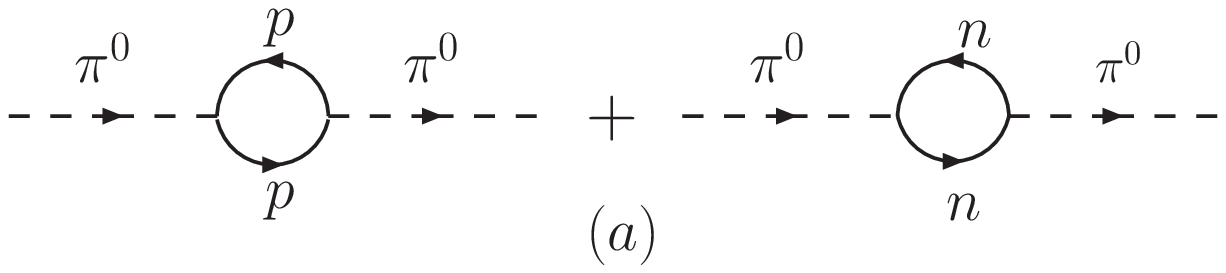, height=2.2 cm} \\
\epsfig{figure=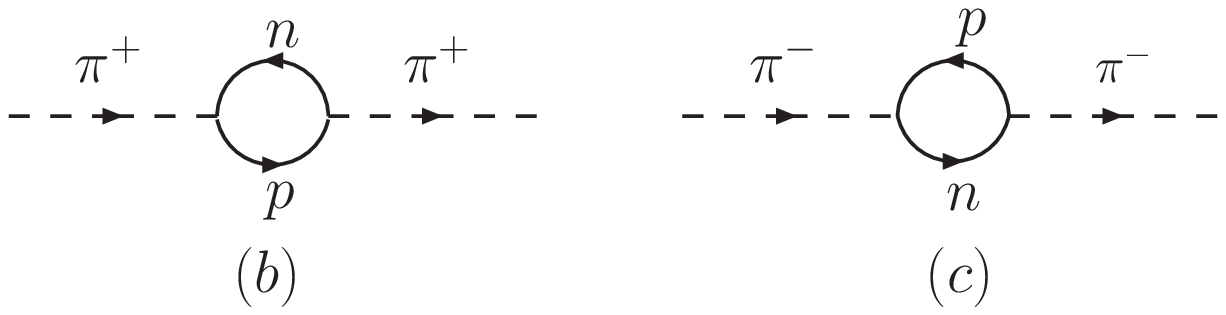, height=2.5cm} \\
\caption{(a) represents the one-loop self-energy diagram for $\p^0$,
  (b) and (c) represents the same for $\p^+$ and $\p^-$ respectively.}
\label{fig1}
\end{center}
\end{figure}
  
Now the self-eneries for $\p^0$ and $\p^\pm$ can be determined by evaluating the 
Fig. 1(a), 1(b) and 1(c) respectively. Using Eqs.\ref{self2} and \ref{self3} in Eq.\ref{self1} 
we get, 

\bea
\S^0_{\p\p}(q) = -~8 ~\lt( \f{f_\p}{m_\p}\rt)^2 \int \f{d^3k}{(2\p)^2\ep^*_k}~{\bf A} 
\label{self4}
\eea
 and 

\bea
\S^\pm_{\p\p}(q) = -~8 ~\lt( \f{f_\p}{m_\p}\rt)^2 \int \f{d^3k}{(2\p)^2\ep^*_k}~
\lt[ {\bf A} \mp {\bf B} \rt] 
\label{self5}
\eea
 
Where 
\bea
{\bf A} = \lt[ \f{M^{*2}q^4}{q^4 - 4(k.q)^2} \rt]\lt( \theta_p + \theta_n \rt)
\label{self6}
\eea
 and 
\bea
{\bf B} = \f{1}{2}\lt[1+ \f{M^{*2}q^2}{q^4 - 4(k.q)^2} \rt](k.q)\lt( \theta_p - \theta_n \rt)
\label{self7}
\eea
 
with $\theta_{p,n} = \theta( k_{p,n} - |{\bf k}|)$. The superscripts on
$\S^{0,\pm}$ represent self-energies for various charged states of pion.
We restrict ourselves in the long wave length limit {\em i.e.} when the pion
momentum ($|{\bf q}|$) is small compared to the Fermi momentum($k^F_{p,n}$) of the 
system where the many body effects manifest strongly. 
In this regime the concept of individual scattering breaks down
and particle propagation can be understood in terms of the low momentum
collective excitations of the system. In the short wave-length limit,
{\em viz. } at distance scale much small compared to the inter particle
separation ($1/k^F_{p,n}$), one 
observes that the particle dispersion approaches 
to that of the free propagation. The recognition of the special role 
played by the distant collisions in determining the collective excitations 
permits analytical solutions of the dispersion relations 
\cite{chin77,abhee03}. 

\section{results}

Now we present main results of our calculations. First we evaluate the
pion self-energy restricting ourselves to the
long wave length limit {\em i.e.} we consider excitations near the Fermi
surface due to scattering. In effect, we assume that the loop momentum
$k \sim k^F$ and $q<<k^F$. This permits us to neglect 
$q^4$ 
compared to the term $4(k.q)^2$ from the denominator of both ${\bf A}$ 
and  ${\bf B}$ in Eqs.\ref{self6} and \ref{self7}. Explicitly, after
performing the integration we get,

\bea
\S^0_{\p\p}(q_0,|{\bf q}|) =\f{1}{2}\lt(\f{f_\p M^*}{m_\p \p}\rt)^2
\lt[\ln \vline\f{1+v_p}{1-v_p}\vline - 
c_0 \ln \vline\f{1+\f{v_p}{c_0}}{1-\f{v_p}{c_0}}\vline~\rt] 
 +[p\ra n]
\label{selfpi0} 
\eea
and 
\bea
\S^\pm_{\p\p}(q_0,|{\bf q}|)=\S^0_{\p\p}(q_0,|{\bf q}|)
\mp \dt\S_{\p\p}(q_0,|{\bf q}|)
\label{selfpipm}
\eea 
where 
\bea
\dt\S_{\p\p}(q_0,|{\bf q}|)&=&\lt(\f{f_\p}{m_\p \p}\rt)^2 \lt[ \f{2}{3}k^3_p q_0
 - \f{M^{*2}q^2}{|{\bf q}|}
\lt(\ep^*_p \ln\lt\vert\f{1+\f{v_p}{c_0}}{1-\f{v_p}{c_0}}\rt\vert
- \f{2M^{*2}}{\sq{c^2_0-1}}\tan^{-1}
\f{k_p\sq{c^2_0-1}}{c_0M^*}\rt)\rt]
-[p\ra n]
\label{8th}
\eea
where $c_0 = q_0/|{\bf q}|$ , $v_{p,n}= k_{p,n}/\ep^*_{p,n}$ and 
$\ep^*_{p,n} = \sq{k^2_{p,n} + M^{*2}} $. 
In the limit of $|{\bf q}|
= 0 $, Eq.\ref{selfpi0} and Eq.\ref{selfpipm} reduces to the following
form
\bea
\S^0_{\p\p}(q_0,0,k_{p,n})=\O^2_{\p\p}q^2_0
\label{3a},~~~~~~
\S^\pm_{\p\p}(q_0,0,k_{p,n})=\O^2_{\p\p}q^2_0 \pm
 m_\p\dt\O^2_{\p\p} q_0,
\label{3b}
\eea
where
\bea
\O^2_{\p\p}&=&\lt(\f{f_\p M^*}{m_\p\p} \rt)^2\lt[
\f{1}{3}\lt(\f{k^3_p}{\ep^{*3}_p}+\f{k^3_n}{\ep^{*3}_n}  \rt) + 
\f{1}{5}\lt(\f{k^5_p}{\ep^{*5}_p}+\f{k^5_n}{\ep^{*5}_n} \rt)
\rt]
\label{sigpipi} \\
\dt\O^2_{\p\p} &=&  \f{1}{m_\p} \lt(\f{f_\p M^* }{m_\p\p} \rt)^2
\lt[ \f{2}{5} \lt( {\frac{k^5_p}{M^{*4}}-\frac{k^5_n}{M^{*4}}} \rt)\rt] 
\label{delsigpipi}
\eea
It is to be noted that both $\O^2_{\p\p}$ and $\dt\O^2_{\p\p}$  
are dimensionless quantities. From Eq.~\ref{3b} and \ref{delsigpipi}, 
it is evident
that the asymmetry driven mass splitting is an order $O(k^{F 5}/M^{* 5})$ 
effect therefore does not appear in calculations at leading order in density.

\subsection{Isospin mode splitting of pion}

Once determined, the self-energies can be used to solve the 
Dyson-Schwinger equation to derive the dispersion relations, 
{\em viz.}
\bea
q_0^2-{\bf q}^2-m^2_{\pi^{0\pm}}-\Sigma^{0,\pm}_{\p\p} (q_0,|{\bf q}|) =0
\eea

In ANM $\S^0 \neq \S^+ \neq \S^-$, hence the pion effective masses will split in matter.
The origin of mass splitting can be understood by considering $\pi^{0,\pm}$ 
scatterings from the Fermi surface in different channels. This
would correspond to various cuts of Fig. 1 due to the $G^D$ term
in Eq.\ref{nuclprop}. 
It is clear that for $\pi^\pm$ scattering we have contributions both from the 
direct ($s$) and exchange ($u$) channels.  However for the $\pi^+$ 
in the $s$ channel neutron ($n$) density
contributes while the $u$ channel involves proton ($p$) Fermi sphere.
For the $\pi^-$ self-energy the role of $p$ and $n$ gets reversed.
While for $\pi^0$ self-energy, both for the $s$ and $u$ channel,
the sum of  $p$ and $n$ densities appear.
 
In the static limit ($|{\bf q}|=0$) following expressions for the pion
effective masses ($q_0$) are found.
\bea
m^{*2}_{\p^0} \simeq \f{m^2_{\p^0}}{1-\O^2_{\p\p}}~~~~~~
m^{*2}_{\p^\pm} \simeq
\f{m^2_{\p^\pm}}{1-(\O^2_{\p\p}\pm\dt\O^2_{\p\p})}
\label{emasses}
\eea
The approximate mass-shifts of $\p^{0\pm}$ are :
\bea
\Dt m_{\p^0} \simeq \f{m_{\p^0}}{2} \O^2_{\p\p},~~~
\Dt m_{\p^\pm} \simeq \f{m_{\p^\pm}}{2}\lt[ \O^2_{\p\p} \pm
\dt\O^2_{\p\p}\rt] \label{mshifts}
\eea
\vskip 0.5 cm
\begin{figure}[t]
\begin{center}
\epsfig{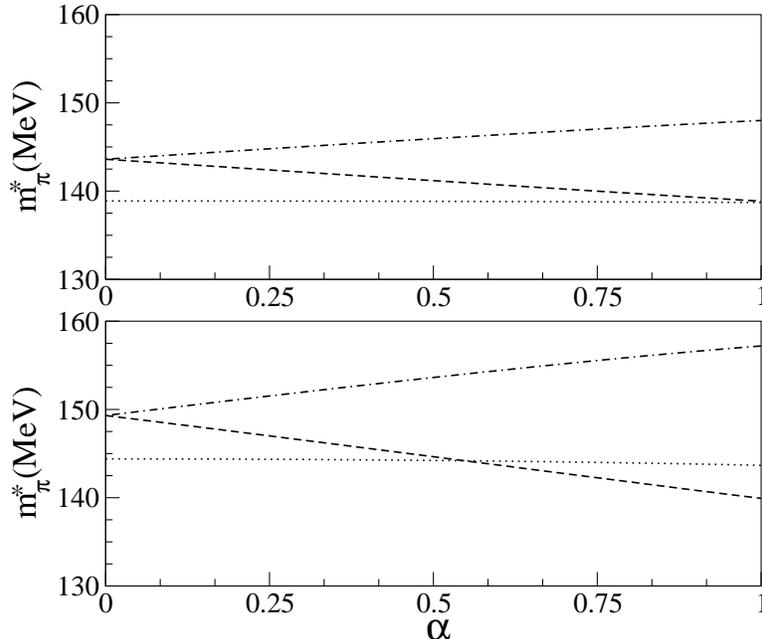} 
\caption{The dashed, dotted and dashed-dotted lines represent the effective masses
  of $\p^+$,$\p^0$,$\p^-$ respectively. Upper panel represents
  effective masses in the free Fermi gas approximation and lower panel 
represents the same with interaction via the scalar mean field.}
\label{fig2}
\end{center}
\end{figure}
From the above equation we find that the splitting is caused by the term 
$\dt\O^2_{\p\p}$ which for SNM ($k_p=k_n$)
vanishes and  $\pi^\pm$ masses, as expected, become degenerate.

Furthermore, it should be
noted that this splitting is an higher order effect 
$\sim k_{p,n}^5/\ep^5_{p,n}$ and therefore at the linear density level
($k_{p,n}^3/\ep^3_{p,n}$) all the pion masses will remain degenerate,
if the free pion masses are assumed to be so.
We have seen that all higher order density corrections to the self-energy
vanish in the static limit $|{\bf q}|=0$.
Numerical estimation for the effective pion masses are shown in Fig.~2
as a function of the asymmetry parameter $\alpha=
(\rho_n-\rho_p)/\rho_b$ at normal nuclear matter density.
Evidently in neutron rich  matter the $\pi^-$ 
meson shows more repulsive behaviour than
$\pi^{0}$ and $\pi^+$ meson which follows from 
Eq. \ref{delsigpipi}. Moreover the $\pi^0$ mass is rather insensitive
to $\alpha$. We find that all the masses receive an overall
density dependent positive corrections consistent with 
Eq.~\ref{sigpipi} and \ref{delsigpipi}. In the upper panel we show
the results corresponding to the free Fermi gas model, while in the lower
panel effect of the interacting ground state is taken via Relativistic
Hartree Approximation (RHA)\cite{serot86}. 

The mean field, as we see from the lower panel of Fig.~2, enhances the
mass splitting. It is observed that although the mass splitting contribute at 
$O(k_{p,n}^5/\ep_{p,n}^5)$, the correction is large compared to the 
vacuum mass difference (Fig.~2). At normal nuclear matter density 
$\Dt m_{\p^+} = -1.15(3.39)$ MeV, $\Dt m_{\p^0} = 4.85(5.01)$ MeV and 
$\Dt m_{\p^-} = 11.19(6.98)$ MeV at $\a = 0.7(0.2)$ with mean field
($M^*_p = M^*_n = 685.0$ MeV).
 
The dispersion relation for $\p^{0\pm}$ is 

\bea
 (q_0^2)_{\p^{0\pm}} \simeq m^{*2}_{\p^{0\pm}}+\g_{\p\p}{\bf q}^2+
\lt( \f{\g^2_{\p\p}}{4}+\a_{\p\p}\rt)\f{{\bf q}^4}{m^{*2}_{\p^{0\pm}}}
\label{disppi0}
\eea

 Effective pion masses $m^*_{\p^{0,\pm}}$ are given by
Eq. \ref{emasses}. Here

\bea
\g_{\p\p}= 1-\chi_{\p\p}+\b_{\p\p}
\label{28a},~~~~~
\chi_{\p\p} = \f{\O^2_{\p\p}}{1-\O^2_{\p\p}}\label{28b}
\eea

\bea
\a_{\p\p} &=& \lt(\f{f_\p M^*}{m_\p\p} \rt)^2 
\f{1}{3(1-\O^2_{\p\p})}\lt(\f{k^3_p}{\ep^{*3}_p}+
\f{k^3_n}{\ep^{*3}_n}  \rt)
\label{28c} \\
\b_{\p\p} &=& \lt(\f{f_\p M^*}{m_\p\p} \rt)^2 
\f{1}{5(1-\O^2_{\p\p})}\lt(\f{k^5_p}{\ep^{*5}_p}+
\f{k^5_n}{\ep^{*5}_n}  \rt)
\label{28d} 
\eea

\subsection{Matter induced $\pi$-$\eta$ mixing }

 Another consequence of $n\leftrightarrow p$
asymmetric ground state, as mentioned before, is the isospin mode mixing.
This is an exclusive density dependent effect. To see this we recall that 

\bea
{\cal L}^{PV}_{\p^0} = - 
\f{ f_\p}{m_\p} \lt[ \bar{\Psi}_p\g_5\g_\m\pr^\m \Phi_{\p^0}\Psi_p -
\bar{\Psi}_n\g_5\g_\m\pr^\m \Phi_{\p}\Psi_n \rt] 
\label{Lpieta1}\\
{\cal L}^{PV}_{\eta} = -
\f{ f_\eta}{m_\eta} \lt[ \bar{\Psi}_p\g_5\g_\m\pr^\m \Phi_{\eta}\Psi_p +
\bar{\Psi}_n\g_5\g_\m\pr^\m \Phi_{\eta}\Psi_n \rt]
\label{Lpieta2}
\eea

 The $\pi^0$ couples to $p$ and $n$ with opposite sign
(Eq.\ref{Lpieta1}) while $\eta$ couples with the same sign (Eq.\ref{Lpieta2}). 
This brings in a relative sign between the proton and neutron loop
as shown in Fig~4 which forces mixing amplitude to vanish in SNM ($k_p=k_n$)
and become non-zero in ANM ($k_p\neq k_n$). The
$\p^0$-$\eta$ mixing amplitude in the static limit is evaluated to be
\bea
\S^0_{\p\eta}(q_0,0)&=&\O^2_{\p\eta}q^2_0
\label{21th}
\eea
where
\bea
\O^2_{\p\eta}&=& \lt(\f{f_\p M^*}{m_\p\p} \rt)
\lt(\f{f_\eta M^*}{m_\eta\p} \rt) \lt[
\f{1}{3}\lt(\f{k^{3}_p}{\ep^{*3}_p}-\f{k^{3}_n}{\ep^{*3}_n}  \rt) + 
\f{1}{5}\lt(\f{k^{5}_p}{\ep^{*5}_p}-\f{k^{5}_n}{\ep^{*5}_n} \rt)
\rt]
\label{22th}
\eea 
Note the difference of sign in Eq.\ref{22th} and Eq.\ref{sigpipi}.  
Clearly $\O^2_{\p\eta}$ is non-zero in ANM and vanishes for $\ep^*_p=\ep^*_n$.

\begin{figure}[tbh]
\begin{center}
\epsfig{figure=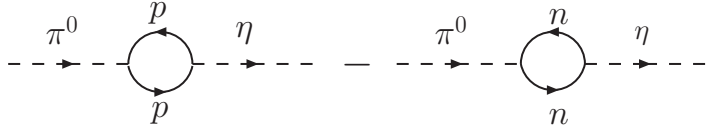, height=2 cm}
\caption{This is the one-loop diagram for $\p^0$-$\eta$ mixing .}
\label{fig4}
\end{center}
\end{figure}

In presence of mixing the pion and eta propagation gets coupled with
each other and can be represented by $2\times2$ matrix 
\cite{serot86,abhee97}
\be
\pmatrix{
1-\f{\S^0_{\p\p}}{q^2_0-m^2_\p} &  \f{\S^0_{\p\eta}}{q^2_0-m^2_\p}\cr
\f{\S^0_{\p\eta}}{q^2_0-m^2_\eta} & 1-\f{\S^0_{\eta\eta}}{q^2_0-m^2_\eta}
} .
\ee

Shifted masses are obtained by solving the following equation.
\bea
(q^2_0-m^2_\p-\S^0_{\p\p}(q_0,0))(q^2_0-m^2_\eta-\S^0_{\eta\eta}(q_0,0))
- (\S^0_{\p\eta}(q_0,0))^2= 0
\label{23th}
\eea

The Eq.\ref{23th} can be written as

\bea
q^4_0(1-\Dt^2)-q^2_0(m^{*2}_{\p^0}+m^{*2}_\eta)+ 
m^{*2}_{\p^0}m^{*2}_\eta = 0 
\label{24th}
\eea

Where $m^*_{\p^0}$ and $m^*_\eta$ are the effective masses 
of $\p^0$ and $\eta$ . Here 

\bea
m^{*2}_{\eta}=\f{m^2_{\eta}}{1-\O^2_{\eta\eta}}~~~~~
\Dt^2=\f{\O^2_{\p\eta}}{1-\O^2_{\p\p}}~
\f{\O^2_{\p\eta}}{1-\O^2_{\eta\eta}}
\label{dispmixpi}
\eea

Solving Eq.\ref{24th} we get the modified effective masses of $\p^0$
and $\eta$ with mixing :
\bea
\tilde{m}_{\p^0}&\simeq& m^*_{\p^0}(1+\f{1}{2}\Dt^2) - \f{m^*_\eta}{2}
\lt[\f{m^*_{\p^0} m^*_\eta }{m^{*2}_\eta -m^{*2}_\p}  \rt]\Dt^2 
\label{26th} \\
\tilde{m}_{\eta}&\simeq& m^*_\eta(1+\f{1}{2}\Dt^2) + \f{m^*_{\p^0}}{2}
\lt[\f{m^*_{\p^0} m^*_\eta }{m^{*2}_\eta -m^{*2}_\p}  \rt]\Dt^2 
\label{27th} 
\eea
Note that for SNM, $\Delta^2=0$ yielding medium modified masses as given
in Eq.\ref{emasses}.

Finally we present pionic dispersion relations in asymmetric nuclear
matter with the possible $\pi^0$-$\eta$ mixing :

\bea
(q^2_0)_{\p^0} &\simeq& \tilde{m}^2_{\p^0}
-\lt[\f{m^{*2}_{\p^0}(\g_{\p\p}+\g_{\eta\eta}-\dt^2)}
{(1-\Dt^2)(m^{*2}_\eta - m^{*2}_{\p^0})} 
-\f{(\g_{\p\p} m^{*2}_\eta + \g_{\eta\eta} m^{*2}_{\p^0})}
{(m^{*2}_\eta - m^{*2}_{\p^0})} \rt]{\bf q}^2 
\label{29a}
\eea

\bea
(q^2_0)_{\eta} &\simeq& \tilde{m}^2_{\eta}
+\lt[ \f{m^{*2}_{\eta}(\g_{\p\p}+\g_{\eta\eta}-\dt^2)}
{(1-\Dt^2)(m^{*2}_\eta - m^{*2}_{\p^0})} 
-\f{(\g_{\p\p} m^{*2}_\eta + \g_{\eta\eta} m^{*2}_{\p^0})}
{(m^{*2}_\eta - m^{*2}_{\p^0})} \rt]{\bf q}^2 
\label{29b}
\eea
Where 
\bea
\dt^2 = 4~\f{\O^2_{\p\eta}}{1-\O^2_{\p\p}}~\lt[
\f{\O^2_{\p\eta}-\f{1}{2}\b_{\p\eta} }{1-\O^2_{\eta\eta}} \rt]
\label{29c}
\eea
\vskip 0.3 cm
\begin{figure}[*htb]
\begin{center}
\epsfig{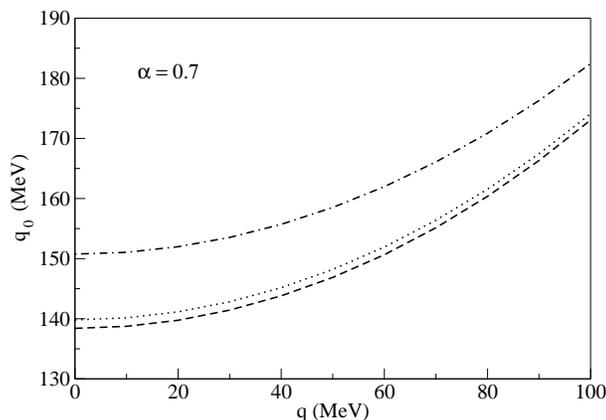} 
\caption{The dashed, dotted and dashed-dotted curves represents the
  dispersion characteristics for $\p^+$,$\p^0$,$\p^-$ respectively at the
  asymmetry parameter $\a = 0.7$.}
\label{fig3}
\end{center}
\end{figure}

In SNM $\Dt^2$ and $\dt^2$ vanish and Eq.\ref{29a}
gives the same dispersion for $\pi^0$ as Eq.\ref{disppi0}. In Fig.~4
dispersion relations for various charged states are depicted.  
The pure density dependent mixing at the
$\eta$ pole is estimated to be $\S_{\p\eta}^{matt}=-1217.475$ MeV$^2$,
 $\S_{\p\eta}^{matt}=-1661.11$ MeV$^2$,
at $\a = 0.2$ and $\a=0.3$ respectively at normal nuclear matter
density with coupling parameters same as that of \cite{piekarewicz93}
which shows that even at normal nuclear matter density the
mixing amplitudes are of the same order as that of the vacuum mixing 
$\S_{\p\eta}^{vac}=-4200$ MeV$^2$ amplitude 
\cite{piekarewicz93}. At higher density the matter induced mixing 
significantly increases. 

\section{Conclusion}

To conclude, in the present work we have shown the characteristic changes
of pion dispersion relations in ANM with the possibility of splitting 
and mixing of different isospin states. A density expansion of the pion
self-energy due to scattering from Fermi sphere is made to obtain analytical 
results. We also have considered the modification of $\pi^0$ mass due to
its mixing with the $\eta$ meson. Furthermore, it is found that the $\pi^-$ in 
neutron rich matter experiences more repulsion than $\pi^{0,+}$ in agreement
with the chiral perturbation theory calculation \cite{weise01}. 
Both the mass splitting (although shown to be an 
$O(k_{n,p}^5/\ep_{n,p}^5)$ effect) and the $\pi$-$\eta$ mixing amplitude
 for neutron rich system are comparable to their vacuum counterparts.
It is found that the density dependent mixing is quite important an effect
even at normal nuclear density for asymmetry in the range of $\a=0.2-0.3$,
at higher density (and/or higher asymmetry ) both the mass 
splitting and mixing effects are dramatically higher.
To focus exclusively on the density 
dependent effects the $\pi$-mass splitting due to $n$-$p$ mass difference has
purposely been neglected. 
Present calculation can be extended to include
such effects which would modify the quantitative results only. 
We also
have included scalar mean field to understand the effect of interacting
ground state.

{\bf Acknowledgment :} Stimulating discussions with Binayak Dutta-Roy is 
gratefully acknowledged. We would also like to thank Jan-e Alam and Pradip Roy 
for carefully reading the manuscript.

\end{document}